\documentclass[namedreferences,hyperref,optionalrh,solaromanenum]{spr-sola}
\usepackage{amsmath,amssymb}
\usepackage{natbib}         
\usepackage{fix-cm}

\usepackage{graphicx}                    
\usepackage{color}                       

\chardef\us=`\_

\newcommand{\revsnd}[1]{}

\begin{document}

\begin{frontmatter}

\title{Modelling the Solar Cycle Nonlinearities into the Algebraic Approach}

%
\author[addressref={aff1},corref,email={mtalafha901@gmail.com}]{\inits{M. H.}\fnm{Mohammed H.}\snm{Talafha}\orcid{0000-0003-4671-1759}}

%
\runningauthor{M. H. Talafha}
\runningtitle{Nonlinear Quenching in Algebraic Method}

\address[id=aff1]{Research Institute of Science and Engineering, University of Sharjah, Sharjah, UAE}

\begin{abstract}
Understanding and predicting solar-cycle variability requires accounting for nonlinear feedbacks that regulate the buildup of the Sun’s polar magnetic field. We present a simplified but physically grounded algebraic approach that models the dipole contribution of active regions (ARs) while incorporating two key nonlinearities: tilt quenching (TQ) and latitude quenching (LQ). Using ensembles of synthetic cycles across the dynamo effectivity range $\lambda_R$, we quantify how these mechanisms suppress the axial dipole and impose self-limiting feedback.  

Our results show that (i) both TQ and LQ reduce the polar field, and together they generate a clear saturation (“ceiling”) of dipole growth with increasing cycle amplitude; (ii) the balance between LQ and TQ, expressed as $R(\lambda_R) = \mathrm{dev(LQ)}/\mathrm{dev(TQ)}$, transitions near $\lambda_R \approx 12^\circ$, with LQ dominating at low $\lambda_R$ and TQ at high $\lambda_R$; (iii) over $8^\circ \leq \lambda_R \leq 20^\circ$, the ratio follows a shallow offset power law with exponent $n \approx 0.36 \pm 0.04$, significantly flatter than the $n=2$ scaling assumed in many surface flux--transport (SFT) models; and (iv) symmetric, tilt-asymmetric, and morphology-asymmetric AR prescriptions yield nearly identical $R(\lambda_R)$ curves, indicating weak sensitivity to AR geometry for fixed transport.  

These findings demonstrate that nonlinear saturation of the solar cycle can be captured efficiently with algebraic formulations, providing a transparent complement to full SFT simulations. The method highlights that the LQ--TQ balance is primarily controlled by transport ($\lambda_R$), not by active-region configuration, and statistically disfavors the SFT-based $1/\lambda_R^{2}$ dependence.
\end{abstract}
%
\keywords{Magnetic fields, Models; Magnetic fields, Photosphere; Magnetohydrodynamics; Solar Cycle, Models.}
\end{frontmatter}

%
\section{Introduction}\label{sect:intro}


The solar dynamo mechanism is a fundamental process responsible for generating the Sun's magnetic field, which in turn drives solar magnetic activity \citep{jiang2016drives}. This mechanism is crucial for understanding the periodic variations in solar phenomena, such as the 11-year sunspot cycle and the periodic reversal of the Sun's magnetic poles

The Sun rotates differentially, meaning the equator rotates faster than the poles. This differential rotation stretches and twists the magnetic field lines, converting the poloidal magnetic field into a toroidal magnetic field \citep{sekii2015recent}.
\revsnd{The Sun’s meridional circulation \--- a large scale poleward flow near the surface with an equatorward return flow at depth \--- transports magnetic flux between low and high latitudes, thereby linking the regions of toroidal- and poloidal-field generation and helping to close the dynamo loop \citep{nandy2004meridional}.}



\revsnd{Although other viable mechanisms \--- such as the turbulent $\alpha$-effect arising from helical convective motions \--- may also contribute to regenerating the Sun’s poloidal magnetic field, the Babcock–Leighton (BL) mechanism remains the leading process. Originally formulated by Babcock and Leighton and revived in modern form by \citet{wang1991magnetic, wang1991new}, it explains the generation of the solar poloidal field through the decay and dispersal of tilted Bipolar Magnetic Regions (BMRs) \citep{choudhuri2018flux, kitchatinov2012solar, jha2020magnetic, choudhuri2023emergence}.} This process involves the emergence and evolution of these tilted BMRs, which contribute to the large-scale dipolar field of the Sun \citep{pal2024algebraic}. The mechanism is a central component of the flux transport dynamo model, where it works in conjunction with the meridional circulation to sustain the solar cycle \citep{hazra2016proposed}. {Additionally, BL} mechanism helps regulate the growth of the Sun's magnetic field. Stronger solar cycles produce BMRs at higher latitudes, which generate less poloidal field, thereby limiting the overall magnetic field growth \citep{karak2020dynamo}. Fluctuations in the {BL} mechanism, along with variations in meridional circulation, are likely causes of the irregularities observed in the solar cycle. These irregularities can be predicted by incorporating observational data into theoretical dynamo models \citep{choudhuri2018flux}.  {Because of this,} the {BL} mechanism is integral to physics-based solar cycle forecasts; quantifying the contribution of individual active regions to the Sun's dipole moment aids in predicting the strength of future solar cycles \revsnd{a correlation first recognised by \cite{schatten1978using} and later refined in subsequent studies} \citep{petrovay2020solar,bhowmik2023physical}. Even during periods of low solar activity, such as grand minima, the {BL} process can still operate with minimal sunspots, allowing the Sun to recover from these inactive phases without requiring an additional generation mechanism for the poloidal field \citep{karak2018recovery}.

The solar magnetic activity cycle exhibits variability within a limited range, indicating the presence of nonlinear mechanisms that prevent the magnetic field's runaway growth \citep{jiang2020nonlinear}. {Without such regulation,} the magnetic field could grow uncontrollably, leading to unrealistic model predictions. Nonlinear feedback mechanisms, such as Latitudinal quenching (LQ) and Tilt quenching (TQ), are crucial \citep{talafha2022role}. 
Similarly, tilt quenching reduces the tilt angles of BMRs, further preventing excessive growth \citep{karak2017solar}. Stronger evidence for LQ compared to TQ, with historical observations {indicates} that sunspot emergence latitudes are higher during stronger cycles \citep{yeates2025latitude}. The dynamo effectivity range, which is influenced by the ratio of equatorial flow divergence to diffusivity, also supports the significance of LQ in modulating the solar cycle \citep{talafha2022role}. Near-surface inflows towards active regions (AR) provide another nonlinear feedback mechanism. These inflows modulate the build-up of polar fields by reducing the tilt angles of BMRs and affecting the cross-equator transport of magnetic flux \citep{talafha2025effect}. {As a result, there is} a strong correlation between the simulated axial dipole strength and the observed amplitude of subsequent cycles, aligning with empirical observations \citep{cameron2012strengths, martin2017inflows, nagy2020impact}. 
\citep{sood2014detailed, kleeorin2023nonlinear}.

Understanding how nonlinearities in AR emergence modulate the solar polar field is crucial for constraining solar cycle variability and predicting future activity. {Although} fully dynamic Surface Flux Transport models (SFT) incorporate these effects through boundary condition adjustments or feedback loops, an algebraic approach offers a simplified, yet physically meaningful, framework to explore their cumulative impact. {In this work, we extend the algebraic approach to quantify the interplay between TQ and LQ systematically. Specifically, by analysing ensembles across the dynamo effectivity range $\lambda_R$, we demonstrate (i) how combined nonlinearities impose a saturation ``ceiling'' on polar field growth, (ii) how the relative dominance of LQ vs. TQ transitions near $\lambda_R \approx 12^{\circ}$, and (iii) that the effective scaling with $\lambda_R$ is much shallower (n $\approx$ 0.36) than the SFT-based n = 2. To ensure the generality of these results, we also test the robustness of these findings under tilt and morphological asymmetries of active regions.}


This paper begins by introducing the algebraic solar dynamo approach and detailing the integration of quenching nonlinearities in Section~\ref{sect:intro}. {Following this, } it outlines the methodology in Section~\ref{sect:method}, including parameter selection, numerical implementation, and simulation setup, {which are} used to investigate the effects of TQ and LQ. The results in Section~\ref{sect:Result} present the simulations' outcomes, {focusing on how} these nonlinearities' impact on the buildup of the solar dipole moment. Subsequently, the discussion in Section~\ref{sect:disc.} interprets the findings in the context of existing literature, evaluates the relative contributions of TQ and LQ to the self-regulation of the solar dynamo, and considers implications for solar cycle prediction. Finally, the conclusion in Section~\ref{sect:concl} summarises the key insights gained from the study.

\section{Methodology} \label{sect:method}

Using the algebraic approach, this study employs a synthetic active region model to investigate the impact of TQ and LQ on the solar axial dipole moment. We explore the role of cycle amplitude and dynamo effectivity range in modulating the dipole moment through the interaction of these quenching effects. The methodology is divided into several steps, including generating synthetic AR, applying quenching mechanisms, and calculating the dipole moment. We then examine the suppression effects and quantify the differences between the quenching modes, No quenching (NoQ), TQ, LQ, and their combination (LQTQ).

\subsection{Synthetic Active Region Cycle Model}

\begin{figure}
	\centering
	\includegraphics[width=0.8\linewidth]{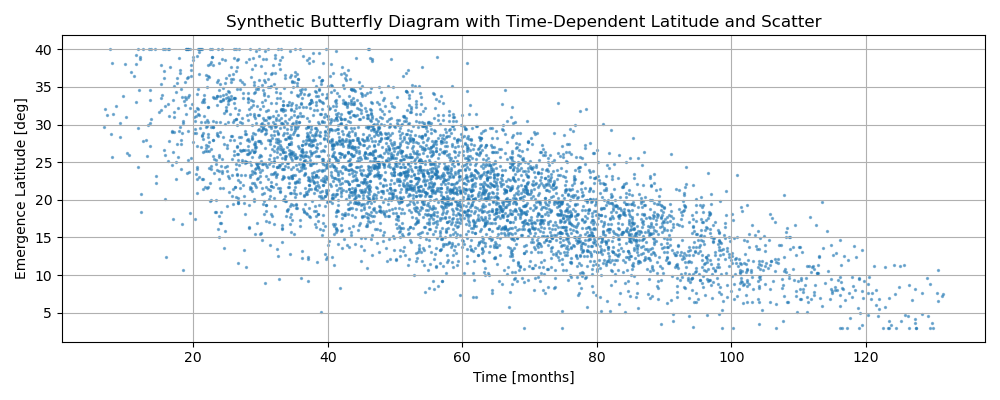}
	\caption{Synthetic butterfly diagram generated using the time-dependent emergence profile from \citet{jiang2018predictability}, with emergence latitudes declining over the cycle and latitude scatter decreasing with time. Each point represents the latitude and time of an individual active region. The distribution captures the equatorward drift and narrowing latitudinal spread of sunspot emergence throughout the solar cycle.}
	
	\label{fig:syn_but}
\end{figure}

To simulate the solar activity cycle, we generate a set of synthetic ARs that emerge throughout a solar cycle. AR properties \--- \revsnd{such as the magnetic flux of each region ($\Phi$) is drawn from a lognormal distribution, while the emergence latitude ($\lambda$) and tilt angle ($\delta$) are drawn from normal distributions following the statistical formulation of \cite{lemerle2015coupled}.} This probabilistic sampling introduces realistic variability and stochasticity in the synthetic AR population, mimicking the observed scatter in solar magnetograms. The latitude distribution captures the butterfly-shaped migration of sunspot groups over the cycle, while the Gaussian tilt distribution around Joy’s law generates diverse polarity separations, enhancing the physical realism of the simulated dipole contributions. Such stochastic modelling also introduces natural fluctuations in the SFT output, reflecting variability in the polar field buildup and allowing a more nuanced analysis of nonlinear dynamo behaviour. In each simulation, the cycle amplitude ratio ($A_n / A_0$) represents the ratio of a cycle amplitude relative to a reference cycle amplitude, and the set of amplitude ratios extends between 1.1 and 3.5 to mimic the solar cycle amplitudes. {We impose a cap of N = 200  candidate ARs per cycle as a numerical sampling choice (not an observed count). The effective emergence rate is governed by the prescribed cycle envelope, and all diagnostics are normalised by total flux; if desired, N can be rescaled to match a given cycle’s sunspot number without changing the results.} The emergence time profile follows Equation~(1) in \cite{jiang2018predictability}, where the shape parameter \( b \) is dynamically adjusted as a function of the cycle amplitude parameter \( a \), according to the relation \( b = 27.12 + 25.15 / (a \times 10^3)^{1/4} \). This {formulation} ensures that stronger cycles exhibit more sharply peaked emergence rates, in line with observed solar cycle asymmetries. The resulting function captures the fast rise and slow decay of solar activity over the 11-year cycle and serves as a basis for sampling the temporal distribution of AR emergence. Figure~\ref{fig:syn_but} illustrates the synthetic butterfly diagram produced using this time-dependent AR emergence equation, demonstrating the equatorward migration and decreasing latitude scatter characteristic of solar cycles.

To generate realistic AR flux values, we adopt a log-normal distribution, consistent with empirical observations of sunspot group and BMR fluxes, with a mean value ($\mu = 22.3$) and standard deviation ($\sigma=0.3$) representing the solar activity level. The logarithm of the flux is sampled from a normal distribution,
\begin{equation}
    \log_{10} \Phi_{\mathrm{rel}} \sim \mathcal{N}(\mu, \sigma)
\end{equation}

and the physical flux is recovered via exponentiation. To scale the AR fluxes according to the strength of the solar cycle, we multiply by a baseline amplitude factor ($A_i$ =0.015) and a cycle-dependent ratio ($A_n/A_0$), yielding the total AR flux,
\begin{equation}
    \Phi = A_i \cdot \left( \frac{A_n}{A_0} \right) \cdot \Phi_{\mathrm{rel}}
\end{equation}

This procedure ensures {not only} a realistic distribution of AR fluxes {but also modulates} their amplitudes across solar cycles of different strengths \citep{baumann2004evolution}.

\begin{figure}
    \centering
    \includegraphics[width=0.8\linewidth]{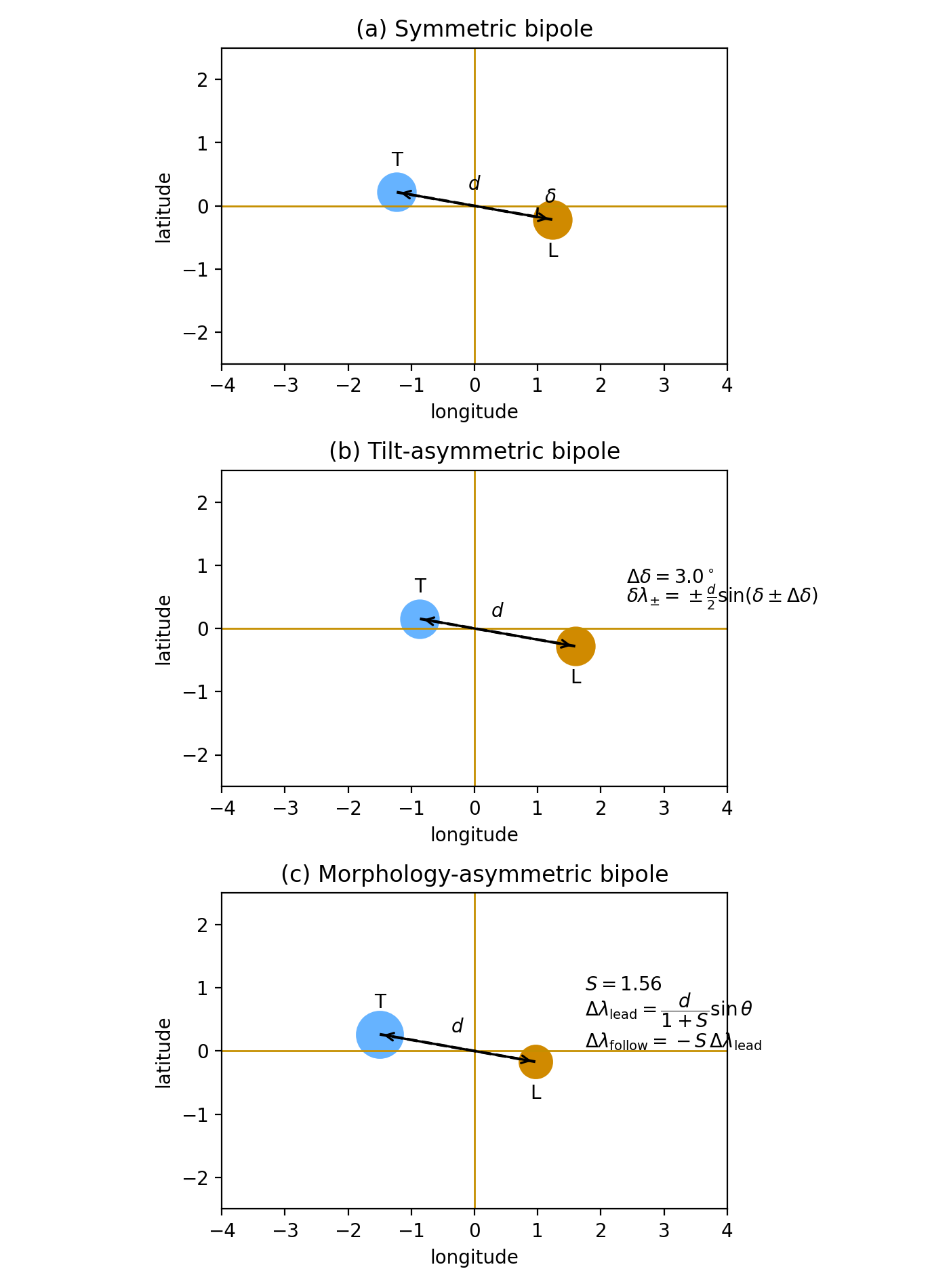}
    \caption{\revsnd{
Schematic of the active-region configurations used in this study. 
Axes denote heliographic longitude and latitude in a local small-angle frame. 
\textbf{(a)}~Symmetric bipole: equal polarities placed symmetrically about the AR centre, with $\delta\lambda_{\pm} = \pm\tfrac{d}{2}\sin\delta$. 
\textbf{(b)}~Tilt-asymmetric bipole: a small angular perturbation $\Delta\delta$ modifies the nominal tilt, producing unequal latitudinal offsets $\delta\lambda_{\pm} = \pm\tfrac{d}{2}\sin(\delta \pm \Delta\delta)$. 
For clarity, $\Delta\delta = 3^{\circ}$ is shown, while in the simulations it is drawn from a normal distribution centred on zero. 
\textbf{(c)}~Morphology-asymmetric bipole: polarities have unequal areas parametrised by the spreading factor $S$. 
The trailing (blue) polarity is shown larger ($S = 1.56$ here), whereas in simulations, $S$ follows a normal distribution consistent with observations. 
All panels share fixed centroids and identical scales.
}}
    \label{fig:sketch}
\end{figure}

The tilt angle \( \delta \) of each AR is assigned according to Joy's Law, following the empirical relation
\begin{equation}
    \delta = \delta_0 \sin \lambda
    \label{eq:joys}
\end{equation}

where \( \lambda \) is the latitude of emergence and \( \delta_0 = 1.5^\circ \) is the proportionality constant. This formulation reflects the observed tendency for tilt angles to increase with latitude and is consistent with previous SFT models, as shown by \cite{wang1991magnetic, baumann2004evolution}. The separation \( d \) between the opposite polarities of each AR is fixed at \( 5^\circ \), consistent with the idealised modelling approach used by \cite{wang2021algebraic}. This value approximates the typical latitudinal span of BMRs and allows for controlled comparisons across quenching scenarios. {While} observational studies suggest that polarity separation increases with magnetic flux \citep{baumann2004evolution}, we {intentionally} adopt a constant \( d \) to isolate the effects of latitude, tilt, and morphological asymmetries on dipole moment generation. {Figure~\ref{fig:sketch} illustrates the active-region prescriptions used in this work, the symmetric bipole (a), tilt-asymmetric bipole (b), and the morphology-asymmetric bipole (c). }

\subsection{Quenching Mechanisms}

To model the suppression of the dipole moment during strong cycles, we implement three quenching mechanisms: TQ, LQ, and their combination (LQTQ). 

Tilt quenching represents a nonlinear modulation of Joy’s Law, whereby the mean tilt angles of ARs are reduced during stronger solar cycles. This effect has been observed in sunspot records \citep{dasi2010sunspot} and interpreted as a dynamical feedback that limits the growth of the solar polar field during high-activity cycles \citep{cameron2010surface, iijima2017improvement}. Following the work of \cite{talafha2022role}, tilt quenching is implemented by scaling the nominal Joy’s Law tilt angle (Equation~\ref{eq:joys}) with a suppression factor that depends on the cycle amplitude. Equation~\ref{eq:joys} is reduced as
\begin{equation}
    \delta_{\mathrm{TQ}} = \delta_0 \sin \lambda \left(1 - b_{\mathrm{joy}} \left(\frac{A_n}{A_0} - 1\right) \right)
    \label{eq:TQ_amp_based}
\end{equation}

where \( b_{\mathrm{joy}} \) is a quenching coefficient set as a constant ($b_{\mathrm{joy}} = 0.15$). {Although the suppression factor depends on the global cycle amplitude as a proxy for the global Lorentz feedback, it is applied to each AR’s tilt. For internal consistency, an equivalent flux-based form can also be written,}
\begin{equation}
\delta_i \;=\; \delta_{\mathrm{Joy}}(\lambda_i)\, f_{\mathrm{TQ}}(\Phi_i),
\qquad
f_{\mathrm{TQ}}(\Phi_i) \;=\; \left[1+\left(\frac{|\Phi_i|}{\Phi_0}\right)^{m}\right]^{-1},
\label{eq:TQ_flux_based}
\end{equation}
{where $\Phi_i$ is the unsigned magnetic flux of AR $i$, $\Phi_0$ is a flux scale, and $m$ controls the nonlinearity. While both parametrisations yield qualitatively similar trends, we retain the amplitude-based form \eqref{eq:TQ_amp_based} to limit free parameters while keeping the quenching operation per-AR.}

Latitude quenching represents that the observed ARs tend to emerge at higher latitudes during stronger solar cycles. Since the axial dipole contribution of an AR decreases with increasing latitude, due to reduced cross-equatorial flux cancellation, this effect acts as a nonlinear saturation mechanism in the solar dynamo. To model latitude quenching, we adjust the mean emergence latitude \( \lambda_0 \) as a function of cycle strength \( A_n / A_0 \):
\begin{equation}
    \lambda_{\mathrm{LQ}} = \lambda_0 + b_{\mathrm{lat}} \left( \frac{A_n}{A_0} - 1 \right)
\end{equation}
where \( b_{\mathrm{lat}} \) is a quenching parameter that controls how strongly the latitude increases with amplitude. {Following} \cite{jiang2011solar} observational analysis, this parameter is set to 2.4. These mechanisms operate jointly to suppress the axial dipole contribution, with TQ limiting the latitudinal separation of polarities and LQ reducing the cross-equatorial flux cancellation efficiency.

\subsection{Dipole Moment Calculation}
To quantify the contribution of ARs to the solar axial dipole moment, two algebraic approaches have been developed based on the SFT model. The first, introduced by \cite{petrovay2020towards}, models each AR as a symmetric bipole and expresses its final dipole contribution as a product of its initial moment and a latitude-dependent Gaussian attenuation factor. While this formulation offers analytic clarity, it assumes an idealised bipolar structure and is not directly applicable to complex or asymmetric ARs. In contrast, \cite{wang2021algebraic} proposed a generalised method that integrates over the full magnetic field distribution of an AR, weighting each flux element by an error function kernel that captures latitude-dependent transport efficiency. This approach retains the linearity of SFT and accommodates arbitrary AR morphologies. Given its broader applicability and closer alignment with observational AR complexity, we adopt \cite{wang2021algebraic} formulation in our analysis. The contribution to the dipole moment from each AR is calculated as:

\begin{equation}
    \delta D_{1} = \sum_{i=1}^{2} \Phi_i \cdot \operatorname{erf} \left( \frac{|\lambda_i|}{\sqrt{2} \lambda_R} \right) \cdot \operatorname{sign}(\lambda_i)
    \label{eq:ar_dipole}
\end{equation}

Here, \( \Phi_i \) is the magnetic flux of the $i$-th polarity (in Mx), \( \lambda_i \) is its heliographic latitude (in degrees), and \( \lambda_R \) is the dynamo effectivity range (in degrees), which extends in range 8$^\circ$ - 20$^\circ$ to mimic the SFT model optimized parameters \citep{petrovay2019optimization}. {The dynamo effectivity range quantifies the balance between advection by the meridional flow and diffusion by turbulent mixing in the surface transport of magnetic flux. Defining the characteristic length scale $L$ (e.g., the pole–equator distance or the active-belt width), the advective and diffusive timescales are}
\[
t_{\rm adv}=\frac{L}{U_0},\qquad t_{\rm diff}=\frac{L^2}{\eta},
\]
{with $U_0$ the surface meridional flow speed and $\eta$ the (supergranular) turbulent diffusivity. A convenient dimensionless measure is therefore}
\begin{equation}
\Lambda \;\equiv\; \frac{t_{\rm diff}}{t_{\rm adv}} \;=\; \frac{U_0\,L}{\eta}\,,
\label{eq:Pe_number}
\end{equation}
\revsnd{which represents the Magnetic Reynolds number for the SFT}. In our notation, Equation~\ref{eq:ar_dipole} uses the reciprocal form
\begin{equation}
\lambda_R \;\equiv\; \frac{\eta}{U_0\,L} \;=\; \Lambda^{-1},
\label{eq:lambdaR_def}
\end{equation}

{under this definition, small $\lambda_R$ indicates advection-dominated transport \--- efficient poleward conveyance of trailing-polarity flux and stronger polar fields \--- whereas large $\lambda_R$ indicates diffusion-dominated transport, enhancing cross-equatorial cancellation and weakening the polar field.}

{The error function {in Equation~\ref{eq:ar_dipole}} serves as a smooth weighting kernel that increases with distance from the equator. This kernel reflects the reduction in cross–equatorial flux cancellation (which yields less efficient dipole build–up). The same kernel is applied consistently across all modes in the simulation.
}

By summing over all ARs in a given cycle, we obtain the total dipole moment contribution under each quenching scenario, enabling quantitative comparisons of the nonlinear suppression effects due to quenching mechanisms. The total dipole moment $D_{\text{tot}}$ for the entire cycle is obtained by summing over the contributions from all active regions:
\begin{equation}
    D_{\text{tot}} = \sum_{i=1}^{N_{\text{AR}}} \delta D_{1,i}
\end{equation}

For each quenching mode, the suppression of the dipole moment is quantified by calculating the deviation from the NoQ case; for example, $\Delta D_{\text{TQ}} = D_{\text{NoQ}} - D_{\text{TQ}}$, these deviations are {then} analysed to quantify the effects of TQ, LQ, and LQTQ on the solar dipole moment across different cycle amplitudes and latitudes.

\section{Results}\label{sect:Result}
\subsection{Nonlinearities in Algebraic Model}
\begin{figure}
    \centering
    \includegraphics[width=0.8\linewidth]{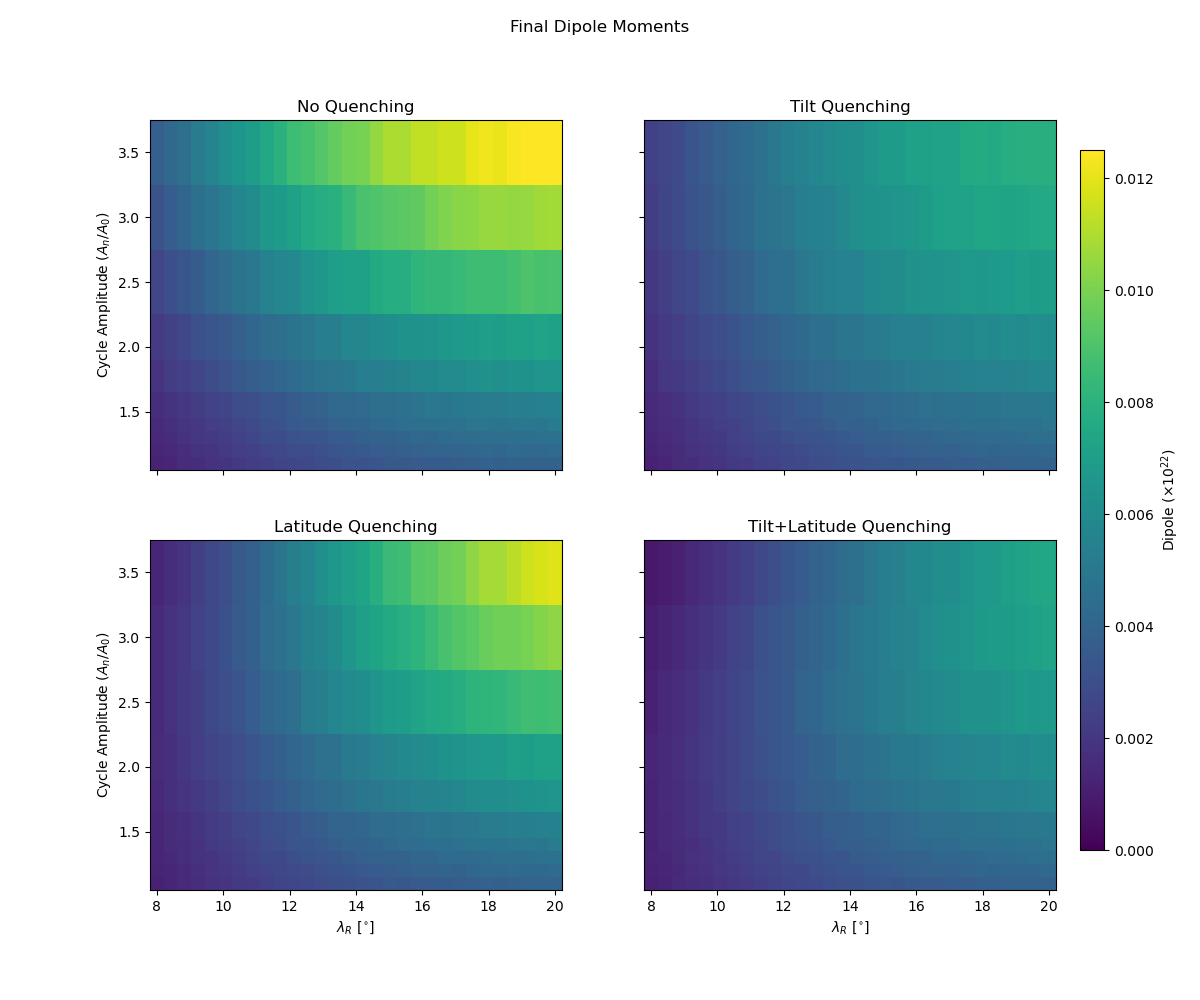}
    \caption{Final axial dipole moments as a function of normalised cycle amplitude ($A_n/A_0$) and dynamo effectivity range ($\lambda_R$) at the end of synthetic solar cycles. Each panel corresponds to a different nonlinear feedback configuration: \textit{Top left} — no quenching; \textit{Top right} — tilt quenching only; \textit{Bottom left} — latitude quenching only; \textit{Bottom right} — both tilt and latitude quenching. Colour indicates the strength of the resulting dipole moment (in units of $10^{22}$ Mx).}
    \label{fig:no_quenching}
\end{figure}

To establish a baseline for comparison, we first analyse the final axial dipole moment generated by 25 synthetic solar cycles in the absence of any nonlinear suppression mechanisms. Figure~\ref{fig:no_quenching} displays a 2D heatmap of the final dipole strength as a function of the cycle amplitude ratio $(A_n/A_0)$ and the dynamo effectivity range $(\lambda_R)$ for four configurations: NoQ (top left), TQ (top right), LQ (bottom left), and both quenching mechanisms LQTQ (bottom right). 

Dipole strength increases with both amplitudes and $\lambda_R$; higher values of \( A_n/A_0 \) correspond to stronger active region emergence rates, directly contributing more flux to the global field, larger values of \( \lambda_R \) represent reduced turbulent diffusion or more extended transport ranges, allowing more efficient poleward migration of trailing polarity flux, which enhances dipole buildup. The relationship is smooth and monotonic, as expected in the absence of feedback mechanisms (top left); for a fixed dynamo effectivity range, increasing \( A_n/A_0 \) leads to a linear rise in the final dipole. However, for a fixed amplitude, increasing \( \lambda_R \) improves transport efficiency, increasing the dipole moment. 

The inclusion of TQ introduces a nonlinear feedback mechanism {in which} the tilt angle of active regions is suppressed as the solar cycle amplitude increases. This effect is based on the premise that stronger toroidal fields reduce the Coriolis-induced tilts of emerging BMRs, thus diminishing their contribution to the global dipole field. As shown in the top right of Figure~\ref{fig:no_quenching}, the final axial dipole moment remains positively correlated with both the cycle amplitude \( A_n/A_0 \) and the dynamo effectivity range \( \lambda_R \). However, the rate of increase is significantly reduced compared to the no quenching scenario. The suppression is most pronounced at high cycle amplitudes, indicating the efficacy of TQ in regulating the buildup of polar fields.

Latitude quenching introduces a nonlinear modulation of AR emergence latitudes with increasing cycle amplitude. As the toroidal field strengthens, ARs tend to emerge at systematically higher latitudes. {Although Joy's law implies the tilt \--- and thus the north–south separation of polarities \--- increases with latitude, the accompanying poleward shift reduces cross–equatorial cancellation of the leading polarity. As a result, the net build–up of the axial dipole becomes less efficient despite the larger tilt.} 

In the bottom left of Figure~\ref{fig:no_quenching}, the final dipole moments under the influence of LQ across a range of cycle amplitude ratios \( A_n/A_0 \) and dynamo effectivity range \( \lambda_R \) are shown. Compared to the NoQ scenario, the overall suppression of the dipole moment is weaker in LQ. The dipole strength rises with increasing \( A_n/A_0 \), especially for larger \( \lambda_R \), indicating that LQ alone does not significantly limit polar field generation. This result is consistent with \cite{jiang2014magnetic} {which found} that LQ has a more subtle regulatory effect than TQ. {Nevertheless}, the saturation trend at high latitudes hints at a nonlinear ceiling in dipole buildup imposed by latitude modulation. {In other words,} while higher cycle amplitudes still lead to more ARs and, in principle, stronger magnetic fields, those ARs are less effective at building up the solar dipole because dipole contribution is strongest at lower latitudes. For very strong cycles (large $A_{n}$) and large $\lambda_R$, the dipole moment no longer increases linearly. It starts to level off or saturate. This means that even if the cycle amplitude increases, the resulting dipole field doesn’t grow proportionally. 

The bottom right of the same figure shows the combined effect of LQ and TQ. In this case, the model {demonstrates} a compounded suppression of the axial dipole moment. Compared to the NoQ, TQ, and LQ cases, the LQTQ configuration produces the most significant reduction in dipole moment, particularly in high-amplitude and high-diffusion regimes. 

The interaction between TQ and LQ {is synergistic rather than purely additive}. TQ reduces the effective tilt angle of active regions, limiting the latitudinal separation of polarities, while LQ shifts the emergence latitude poleward, further diminishing the likelihood of cross-equatorial flux cancellation. {Acting} together, these mechanisms constrain both the geometry and location of AR contributions, thereby curtailing the buildup of the axial dipole field more effectively than either mechanism alone. This collective suppression introduces a natural saturation mechanism in the {BL} dynamo framework. It curbs the growth of polar fields during strong cycles, potentially explaining the observed tendency for the Sun to self-regulate its magnetic output. Furthermore, the nonlinear damping introduced by LQTQ may contribute to the reduction in cycle-to-cycle memory and the limited predictability horizon of solar activity.

\begin{figure}[htp!]
    \centering
    \includegraphics[width=0.8\textwidth]{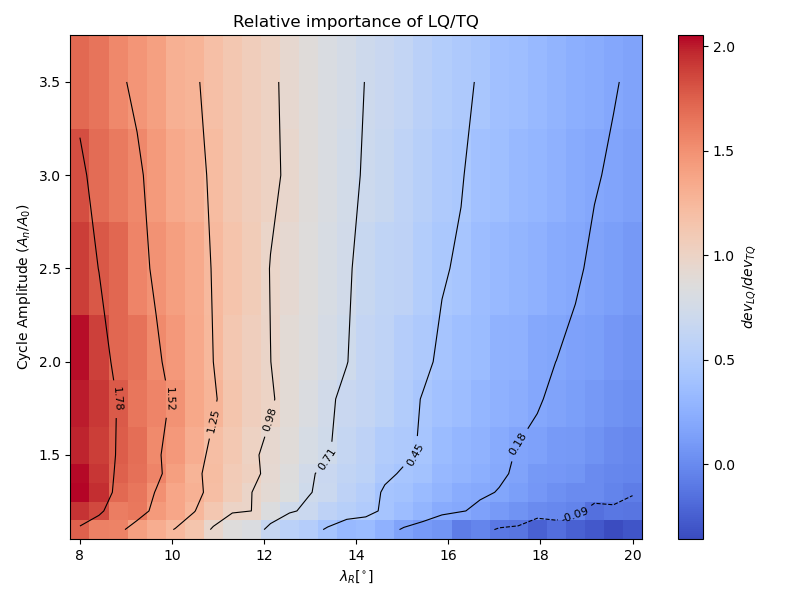}
    \caption{Relative importance of LQ \textit{vs.} TQ in suppressing the solar axial dipole moment, shown as a function of dynamo effectivity range ($\lambda_{R}$) and cycle amplitude ratio ($A_n/A_0$). The colourmap indicates the ratio {R($\lambda_R$)}, with red regions indicating LQ-dominated suppression and blue regions indicating TQ-dominated suppression. Contour lines provide quantitative values of the ratio.}
    \label{fig:LQ_TQ_importance}
\end{figure}

Figure~\ref{fig:LQ_TQ_importance} illustrates the relative importance of LQ versus TQ in suppressing the solar axial dipole moment, as a function of the dynamo effectivity range ($\lambda_{R}$) and cycle amplitude ratios ($A_n/A_0$). The quantity, {R($\lambda_R$)= $\mathrm{dev}_{\mathrm{LQ}} / \mathrm{dev}_{\mathrm{TQ}}$}, represents the ratio of dipole moment suppression from no quenching case due to LQ to that from TQ. Regions shaded in red ({R($\lambda_R$)} $\gtrsim 1$) indicate dominance of LQ, while blue regions ({R($\lambda_R$)} $\lesssim 1$) suggest stronger suppression from TQ. Contour lines overlay the colour map to provide quantitative values of the ratio, delineating the transition zones between the two regimes.

The distribution of red and blue regions reveals that when \(\lambda_{R}\lesssim11^\circ\) (i.e.\ when meridional flow dominates over diffusion) and for strong cycles (\(A_n/A_0\gtrsim1.5\)), {LQ emerges as the primary} nonlinear mechanism, reflecting the tight confinement of flux near the emergence latitude.  As \(\lambda_{R}\) increases, indicating a rising role of diffusion, the efficacy of {LQ} diminishes and {TQ} takes over at higher latitudes (notably beyond \(\sim15^\circ\)), as shown by the blue shading.  The smooth progression of the unity contour demonstrates how modest changes in the diffusivity\--\ flow balance or cycle strength can tip the scale between LQ and TQ.

\begin{figure}[ht]
    \centering
    \includegraphics[width=0.8\textwidth]{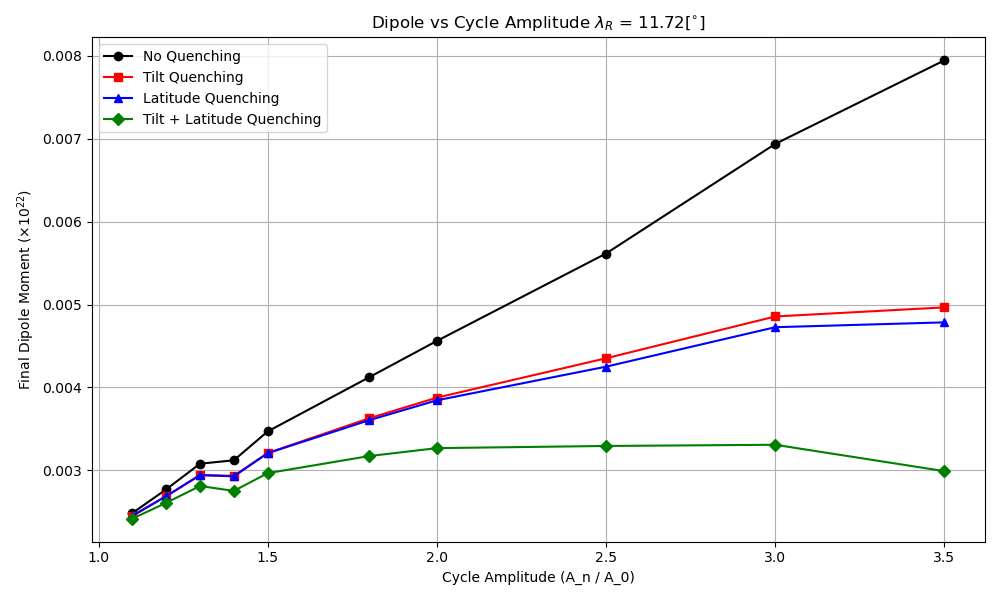}
    \caption{Final axial dipole moment (in units of $10^{22}\,$Mx) at cycle minimum versus cycle amplitude ratio $A_n/A_0$, at $\lambda_R = 11.72^\circ$.  Curves show results with NoQ (black), TQ (red), LQ (blue), and LQTQ (green). } \label{fig:dipole_vs_amplitude}
\end{figure}

As shown in Figure~\ref{fig:dipole_vs_amplitude}, the final axial dipole moment \(D_{\rm final}\) at cycle minimum is plotted against the cycle amplitude ratio \(A_n/A_0\) for four scenarios at $\lambda_R = 11.72^\circ$ as an example. In the absence of any nonlinear feedback, \(D_{\rm final}\) increases linearly with \(A_n/A_0\), reflecting a direct proportionality between the emerging bipole flux and the polar‐field buildup. {When TQ is introduced alone} (\(b_{\rm joy}=0.15\)) {the response becomes nonlinear}: as cycle strength grows, the mean active‑region tilt is reduced, diminishing the net dipole contribution per unit flux. {LQ} alone (\(b_{\rm lat}=2.4\)) likewise yields a nonlinearity, since stronger cycles emerge, on average, at higher latitudes, reducing cross‑equatorial cancellation and thus polar flux generation. {When TQ and LQ act together}, the dipole amplitude exhibits pronounced saturation for \(A_n/A_0\gtrsim2\), and even a slight downturn at the highest amplitudes. This “ceiling” effect demonstrates that the combined action of reduced tilt and shifted emergence latitude imposes an effective self‑limiting feedback on polar‑field growth.

\begin{figure}
    \centering
    \includegraphics[width=0.8\linewidth]{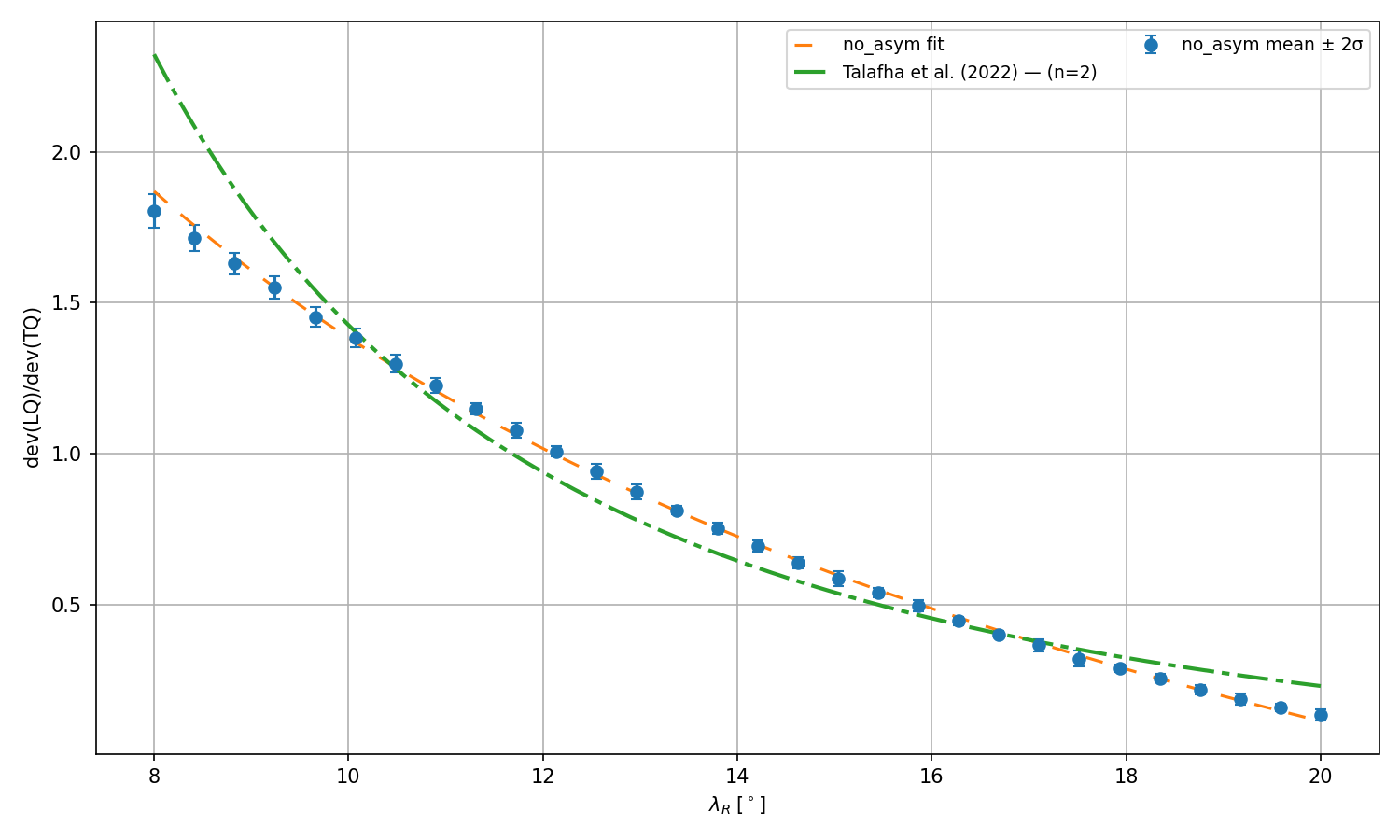}
    \caption{The relative importance of LQ vs. TQ, as a function of the dynamo effectivity range \( \lambda_R \) at fixed cycle amplitude ratio \( A_n/A_0 = 3.0 \), {for Symmetric case. Fitting line follows Equation~\ref{eq:fit}, the vertical error bars show \(\pm2\sigma\), dashed-dot green line represents the fitting from SFT simulations in \cite{talafha2022role} with n=2.}  }

    \label{fig:rel_impor}
\end{figure}

For an arbitrary cycle amplitude ratio \( A_n/A_0 = 3.0 \), we calculated the ratio {R($\lambda_R$)}, Figure~\ref{fig:rel_impor} show the relative importance of LQ \textit{vs.} TQ with respect to all values across $\lambda_R$ range, {for an ensemble of 10 statistically independent simulations. The fit line follows the form:} 
\begin{equation}
    R(\lambda_R) \approx C_1 + \frac{C_2}{\lambda_R^n}
    \label{eq:fit}
\end{equation}

{Where the fitting parameters are listed in Table~\ref{table:fitting_par} for all cases considered in this study.}

\begin{table}
\caption{Best–fit parameters for the ratio $R(\lambda_R)$ obtained by $\chi^2$ minimisation of the ensemble means. Each row corresponds to the algebraic simulations (Symmetric, tilt asymmetry, morphological asymmetry). Each case lists the coefficients $C_1$, $C_2$ and $n$, with their $1\sigma$ uncertainties, \revsnd{together with the reduced chi-square values $\chi_{\nu}^2 = \chi^{2}/(N-\nu)$ normalised by the number of degrees of freedom}. Pointwise uncertainties are the ensemble $1\sigma$ scatter about the mean (not the standard error) and are used as absolute weights in the fit. When $2\sigma$ bars are shown in figures, the fit still uses $1\sigma$ weights.}

\label{table:fitting_par}
\begin{tabular}{lcccc}    
\hline  \\  
	Case& $C_{1}$& $C_{2}$& $n$ & $\chi_{\nu}^{2}$\\
\\
\hline
Symmetric&-4.118$\pm$0.461&13.157$\pm$0.058&0.378$\pm$0.035&1.574\\
Tilt Asymmetry&-4.156$\pm$0.480&13.159$\pm$0.063&0.375$\pm$0.036&1.488\\
Morphological Asymmetry&-4.403$\pm$0.390&13.159$\pm$0.078&0.356$\pm$0.027&2.323\\
\hline
\end{tabular}
\end{table}

\subsection{Tilt Asymmetry}

Active regions may exhibit asymmetry in their tilt configuration, such that the leading and following polarities are not symmetrically positioned about the mean Joy’s Law tilt axis. Such tilt asymmetry has been observed in sunspot group morphologies and is thought to arise from convective buffeting of emerging loops \citep{longcope1996effects}, anchoring depth differences \citep{d1993theoretical}, or intrinsic asymmetries in flux tube emergence \citep{wang2021algebraic}. When tilt angles are reduced due to TQ or scattered through normal distributions, the dipole moment becomes sensitive to random deviations in polarity separation. The dipole contribution from an AR is proportional to the sine of its tilt angle,  making the Gaussian spread around the mean a significant factor in modulating signed dipole strength. For strong cycles subject to TQ, where the mean tilt is suppressed, the contribution from randomly larger tilt angles sampled from the distribution can partially counteract the suppression. Conversely, random tilts with smaller magnitudes can enhance the reduction in dipole formation. This stochasticity introduces dipole noise and reduces the predictability of the final polar field, especially under strong nonlinear modulation, further emphasising the importance of AR-scale variability in the dynamo process. 

In our model, tilt asymmetry is implemented by introducing a small angular offset \( \Delta \delta \) between the effective tilt angles of the two polarities. This modifies the latitudinal separation such that:
\begin{equation}
    \delta \lambda_{\pm} = \pm\frac{d}{2} \sin(\delta \pm \Delta \delta)
\end{equation}
where \( \delta \) is the nominal Joy’s Law tilt and \( d \) is the separation between the polarities.


{Tilt asymmetry is implemented as a per-active-region random perturbation drawn from a zero-mean normal distribution: 
$\Delta\delta \sim \mathcal{N}(0,\sigma_\Delta^2)$ with $\sigma_\Delta = 1^\circ$; 
the two footpoints are then placed with tilts $\delta\pm\Delta\delta$ so that the mean tilt remains $\delta$.
This phenomenological perturbation is intentionally
small compared with the observed scatter of sunspot-group tilts about Joy’s law (standard deviations
of order $10^\circ$–$30^\circ$), which is commonly attributed to convective buffeting and other local
emergence effects; see, e.g., \citep{dasi2010sunspot,jiang2014effects, jiao2021sunspot}.}

\revsnd{This asymmetry perturbs the active-region geometry, placing one polarity closer to the equator than the other and altering the net dipole contribution. When the equatorward polarity is stronger, enhanced cross-equatorial cancellation reduces the axial dipole; when the poleward polarity dominates, the dipole may increase. Introducing tilt asymmetry broadens the distribution of individual active-region dipole contributions, increasing the stochastic variability of the global dipole (Figure~\ref{fig:sketch}b). Because polar field generation depends sensitively on the latitudinal placement and flux balance of emerging regions, even small tilt asymmetries can produce significant variability, despite their mean effect being weaker than that of tilt or latitude quenching.}

\begin{figure}
    \centering
    \includegraphics[width=0.8\linewidth]{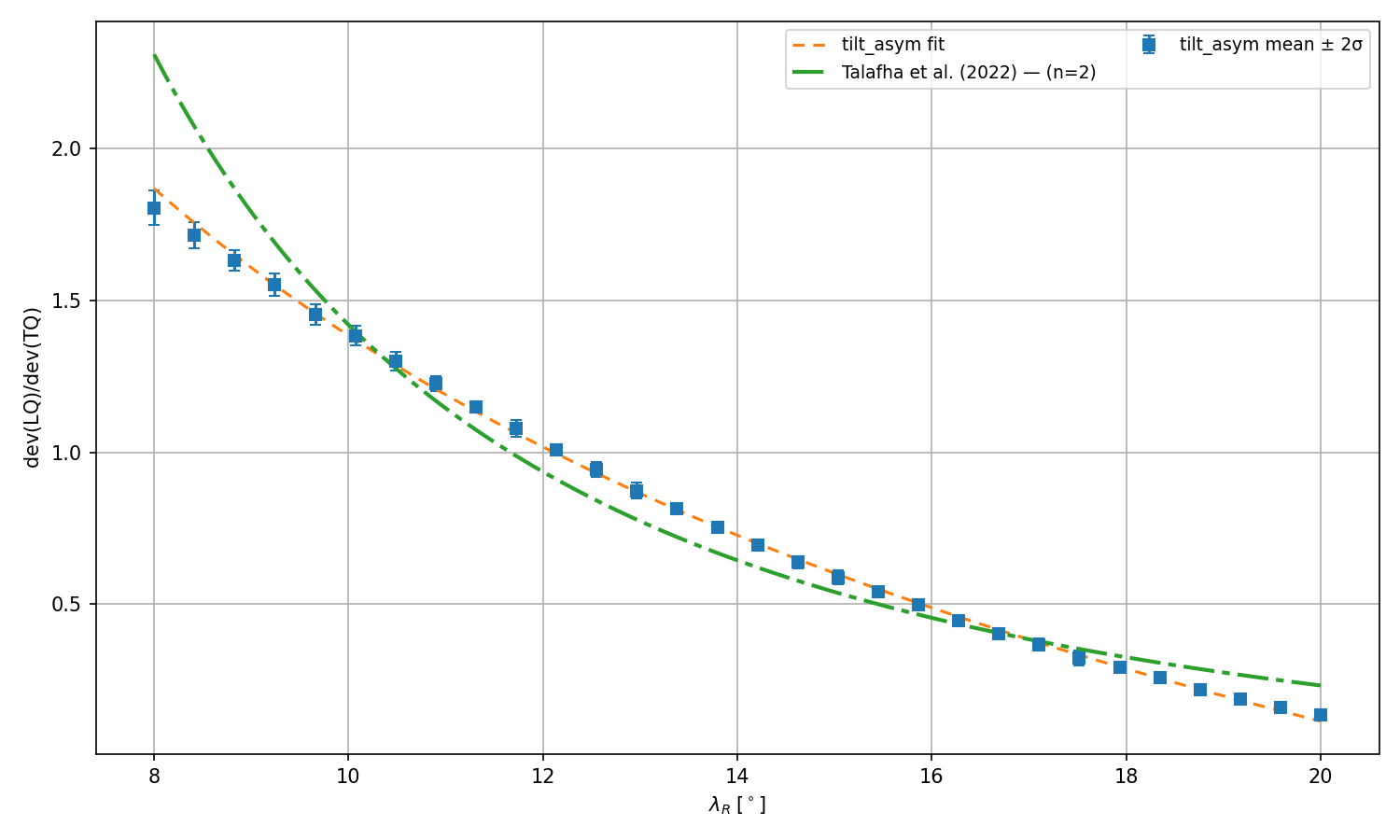}
    \caption{Same as Figure~\ref{fig:rel_impor} including tilt asymmetry to the emerging ARs. }
    \label{fig:rel_impor_TQ}
\end{figure}

Figure~\ref{fig:rel_impor_TQ} shows the relative importance of LQ \textit{vs.} TQ, as a function of the dynamo effectivity range \(\lambda_R\) for a fixed cycle amplitude ratio of \(A_n/A_0 = 3.0\). In this case, the simulation includes hemispheric asymmetry in the tilt angles of emerging bipolar magnetic regions, for 10 ensembles of statistically independent runs, the orange dashed line represents a fit to the ratio $R(\lambda_R)$, fitting parameters are listed in Table~\ref{table:fitting_par}, \revsnd{the goodness of fit was evaluated using the reduced chi-square statistic ($\chi_{\nu}^{2}$), where this normalization by the degrees of freedom ensures that $\chi_{\nu}^{2} \approx 1$ indicates a statistically consistent fit independent of dataset size}. Additionally, the dashed-dot green line show the SFT-based fitting by \cite{talafha2022role}.


To assess the role of tilt asymmetry {using the algebraic approach}, we compare the inclusion tilt asymmetry results with those presented in Figure~\ref{fig:rel_impor}, which presents the same analysis but without imposing any asymmetry in the tilt angles. The qualitative behaviour remains similar: the ratio \(\mathrm{dev}_{\mathrm{LQ}}/\mathrm{dev}_{\mathrm{TQ}}\) decreases monotonically with increasing \(\lambda_R\), again transitioning from LQ-dominated to TQ-dominated. 
The fitted parameters \(C_1 \), \(C_2 \) and \(n\) are nearly identical, \revsnd{confirming the robustness of the shallow power-law scaling obtained from the three-parameter fit}. The comparison indicates that introducing tilt asymmetry slightly enhances the impact of LQ, especially in the diffusive regime, by amplifying latitudinal differences in flux transport and dipole contribution. Thus, while the overall dependency of LQ/TQ importance on \(\lambda_R\) is preserved, \revsnd{introducing tilt asymmetry slightly increases the run-to-run variability, but the ensemble-averaged scaling parameters remain statistically indistinguishable from the symmetric case, confirming that tilt asymmetry does not significantly alter the overall trend in this model.}

\subsection{Morphological Asymmetry}

Observational studies have shown that ARs frequently exhibit morphological imbalances, {with} the leading polarity more compact and concentrated, while the following polarity is more dispersed and diffuse \citep{iijima2019effect,wang2021algebraic}. In this paper, this morphological asymmetry is implemented by introducing a spreading factor to alter the latitudinal separation of polarities while preserving total unsigned flux. Specifically, morphological asymmetry is introduced by assigning a smaller displacement to the leading polarity and a proportionally larger displacement to the following polarity. This is {controlled} by a spreading factor \( S \), such that the latitude shifts become:
\[
\Delta \lambda_{\mathrm{lead}} = \frac{d}{1+S} \sin(\delta), \quad \Delta \lambda_{\mathrm{follow}} = -S \cdot \Delta \lambda_{\mathrm{lead}},
\]
where \( d \) is the base separation and \( \delta \) is the tilt angle. {The factor }\( S \) is randomly drawn from a uniform distribution in the range [1.5, 3.0], reflecting variability in observed AR morphology, {Figure~\ref{fig:sketch}(c)}. This approach preserves the total signed flux but redistributes it spatially; as both polarities emerge from the same subsurface flux loop, allowing us to isolate and quantify the impact of morphological asymmetry on the axial dipole moment buildup under different transport conditions.

\begin{figure}
	\centering
	\includegraphics[width=0.8\linewidth]{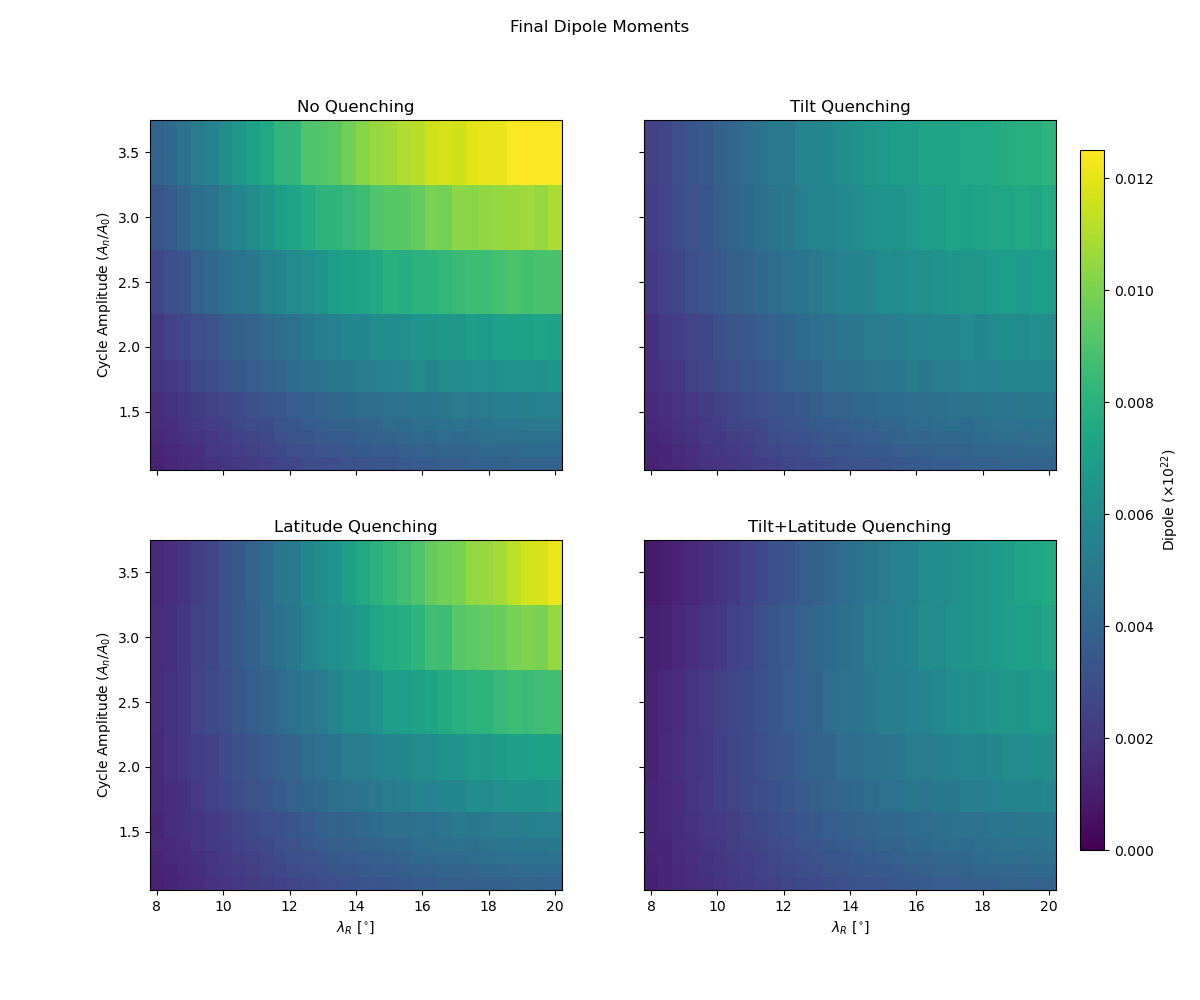}
	\caption{Final axial dipole moments as a function of normalised cycle amplitude ($A_n/A_0$) and dynamo effectivity range ($\lambda_R$) under imposed morphological asymmetry with a random spreading factor.}
	\label{fig:dipole_morph_random}
\end{figure}

Figure~\ref{fig:dipole_morph_random} presents the resulting final dipole moments under the four quenching scenarios. The dipole strength increases systematically with both the cycle amplitude ratio \( A_n/A_0 \) and the dynamo range parameter \( \lambda_R \), consistent with expectations. However, the inclusion of morphological asymmetry modulates this trend. Notably, dipole suppression becomes more pronounced in the LQ and LQTQ panels, especially at low \( \lambda_R \), where the broadened trailing polarity experiences greater diffusive cancellation before reaching the poles.

\begin{figure}[htp!]
	\centering
	\includegraphics[width=0.8\linewidth]{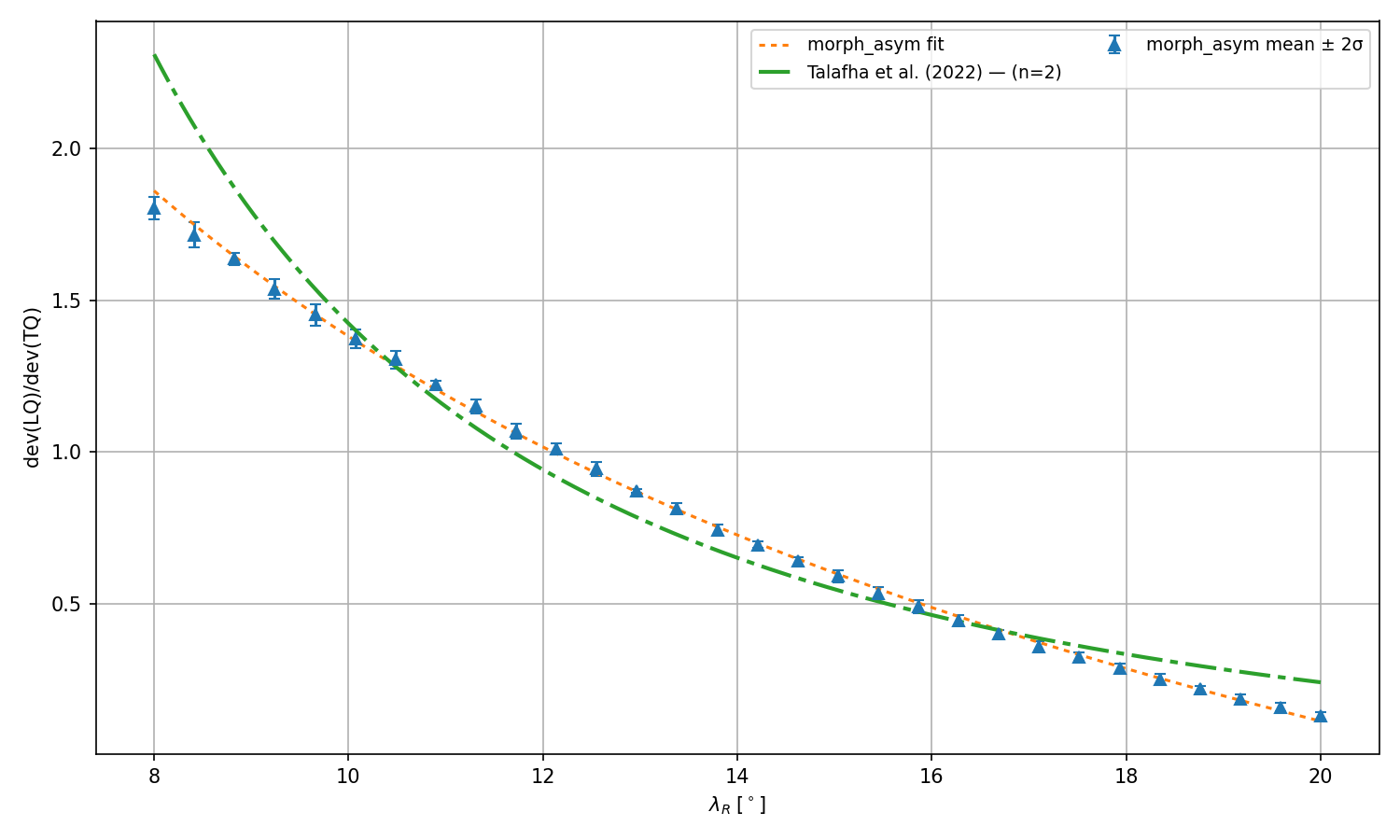}
	\caption{Same as Figure~\ref{fig:rel_impor} including morphological asymmetry with a random spreading factor to the emerging ARs.}
	\label{fig:lq_tq_ratio_random}
\end{figure}

To quantify the relative importance of LQ and TQ under morphological asymmetry, Figure~\ref{fig:lq_tq_ratio_random} shows the ratio {R($\lambda_R$)}, at a fixed cycle amplitude \( A_n/A_0 = 3.0 \). For small values of \( \lambda_R \), this ratio exceeds unity, indicating that LQ dominates due to the enhanced cancellation of dispersed trailing flux. As \( \lambda_R \) increases, the ratio decreases monotonically, and TQ becomes the dominant mechanism beyond \( \lambda_R \sim 13^\circ \). In comparison with no morphological asymmetry, the differences reflect subtle changes in quenching dominance due to randomness in the morphology. The slightly less negative \(C_1\) implies a weaker relative suppression from LQ at high \( \lambda_R \), which is consistent with a reduced spatial dispersion effect.

{It is important to note that the three “asymmetry” variants give nearly identical R($\lambda_R$) because the algebraic model is linear in flux and dominated by latitude and transport; The influence of detailed AR shape enters only weakly, through small factors multiplying $C_{1}$ and $C_2$}

\subsection{SFT-inspired fit (fixed \texorpdfstring{$n=2$}{n=2})}
For comparison with the algebraic-approach results, we perform a single pooled $\chi^2$ fit to all ensemble-mean points with the exponent fixed to $n=2$. Pointwise uncertainties are the ensemble
$1\sigma$ scatter across independent runs and are used as absolute weights in the fit.

The best–fit coefficients and their $1\sigma$ uncertainties are listed in Table~\ref{tab:sft-global}.
This pooled SFT–style model yields a very poor goodness of fit ($\chi^2=8730.7$ for $\nu=88$,
$\chi^2_\nu=99.2$, $p\!\approx\!0$), decisively rejecting a global $n=2$ scaling for our algebraic
ensembles. Nevertheless, the $R(\lambda_R)=1$ crossover occurs at
$\lambda_R^\ast=\sqrt{C_2/(1-C_1)}=11.682\pm0.018^\circ$, consistent with the $\sim12^\circ$
transition seen in the free-$n$ per–case fits. Because the global $n=2$ fit is inadequate, this
formal $1\sigma$ is likely underestimated; inflating by $\sqrt{\chi^2_\nu}$ gives a more realistic
uncertainty of $\pm0.17^\circ$.

\begin{table}[t]
  \centering
  \caption{Best–fit parameters for   $R(\lambda_R)$, obtained by
  $\chi^2$ minimization using the ensemble $1\sigma$ scatter as absolute
  uncertainties. The entry corresponds to the single global curve fit to all
  cases combined. $\lambda_R$ is in degrees; uncertainties are $1\sigma$.}
  \label{tab:sft-global}
  \vspace{2mm}
  \begin{tabular}{lcccc}
    \hline
    case & $c_1$ & $c_2$ & $n$ (fixed) & $\lambda_R^\ast$ (deg) \\
    \hline
    \cite{talafha2022role} &
    $-0.161 \pm 0.002$ &
    $158.454 \pm 0.381$ &
    $2.0$ &
    $11.682$ \\
    \hline
  \end{tabular}
\end{table}

\subsection{Mathematical interpretation}
According to the mathematical interpretation in \cite{talafha2022role} the ratio $R(\lambda_R)$ should follow Equation~\ref{eq:fit} with n=2. In the Algebraic approach, $R$ is evaluated as an ensemble average over the cycle distributions of $(\lambda_0,\alpha,\Phi)$, leading to

\begin{equation}
\big\langle R(\lambda_R)\big\rangle\;\approx\;\langle C_1\rangle+\frac{\langle C_2\rangle}{\lambda_R^2}.
\label{eq:R-avg}
\end{equation}
Over a finite range $\lambda_R$ $\in[8^\circ$,$20^\circ]$, the constant term frequently dominates; consequently, the sum in \eqref{eq:R-avg} is well approximated by a shallow effective power law $C_1+C_2/\lambda_R^{\,n_{\rm eff}}$ with $n_{\rm eff}\ll 2$ (empirically $\sim 0.3$ in our runs).

This can be derived if we let $f(\lambda)\equiv \langle R(\lambda)\rangle=C_1+C_2\lambda^{-2}_{R}$ with $C_1>0$ and consider
the local effective exponent defined on a log--log plot by
\[
n_{\rm eff}(\lambda)\;\equiv\;-\frac{{\rm d}\ln f}{{\rm d}\ln\lambda_R}
\;=\;-\frac{\lambda_R}{f(\lambda_R)}\,\frac{{\rm d}f}{{\rm d}\lambda_R}
\;=\;\frac{2C_2/\lambda^{2}_{R}}{\,C_1+C_2/\lambda^{2}_{R}\,}
\;=\;\frac{2}{\,1+\dfrac{C_1\lambda^{2}_{R}}{C_2}\,}.
\]
Whenever the constant part $C_1$ is comparable to or larger than the decaying part
$C_2\lambda^{-2}_{R}$ on the fitted interval, the ratio $C_1\lambda^2_{R}/C_2\gg 1$ and thus
\[
n_{\rm eff}(\lambda)\;\approx\; \frac{2C_2}{C_1}\,\frac{1}{\lambda^{2}_{R}}\;\ll\;2.
\]
In other words, the presence of the nonzero baseline $C_1$ suppresses the apparent
log--log slope, making $f(\lambda_R)$ look like a much flatter power law over a finite range.
This is precisely the situation in our data on $\lambda_R\in[\lambda_1,\lambda_2]$. A convenient global measure over the finite interval is the secant (average) slope
\[
\bar n\;=\;-\frac{\ln f(\lambda_2)-\ln f(\lambda_1)}{\ln(\lambda_2/\lambda_1)}
\;=\;\frac{\displaystyle \ln\!\Big(1+\frac{C_2}{C_1\lambda_1^{2}}\Big)
-\ln\!\Big(1+\frac{C_2}{C_1\lambda_2^{2}}\Big)}
{\ln(\lambda_2/\lambda_1)}.
\]
If we set $\varepsilon_i\equiv \dfrac{C_2}{C_1\lambda_i^{2}}$ and $\varepsilon_i\ll 1$, then
$\ln(1+\varepsilon_i)\approx \varepsilon_i$ and
\[
\bar n\;\approx\;\frac{\varepsilon_1-\varepsilon_2}{\ln(\lambda_2/\lambda_1)}
\;=\;\frac{C_2}{C_1}\;
\frac{\lambda_1^{-2}-\lambda_2^{-2}}{\ln(\lambda_2/\lambda_1)}
\;\ll\;2.
\]

The formulas above make explicit that
$n_{\rm eff}$ is governed by the dimensionless ratios
$\varepsilon(\lambda)=C_2/(C_1\lambda^{2})$ sampled on the fitted range:
\[
n_{\rm eff}(\lambda)=\frac{2\,\varepsilon(\lambda)}{1+\varepsilon(\lambda)}
\quad\Longrightarrow\quad
\text{for } \varepsilon= \mathcal{O}(10^{-1})\text{ we get } n_{\rm eff}=\mathcal{O}(10^{-1})\ (\approx 0.3).
\]
Thus, a shallow effective exponent is the generic outcome of fitting a constant-plus-decay
with a single offset power over a finite latitude window when the constant term is non-negligible.

\begin{figure}
    \centering
    \includegraphics[width=0.8\linewidth]{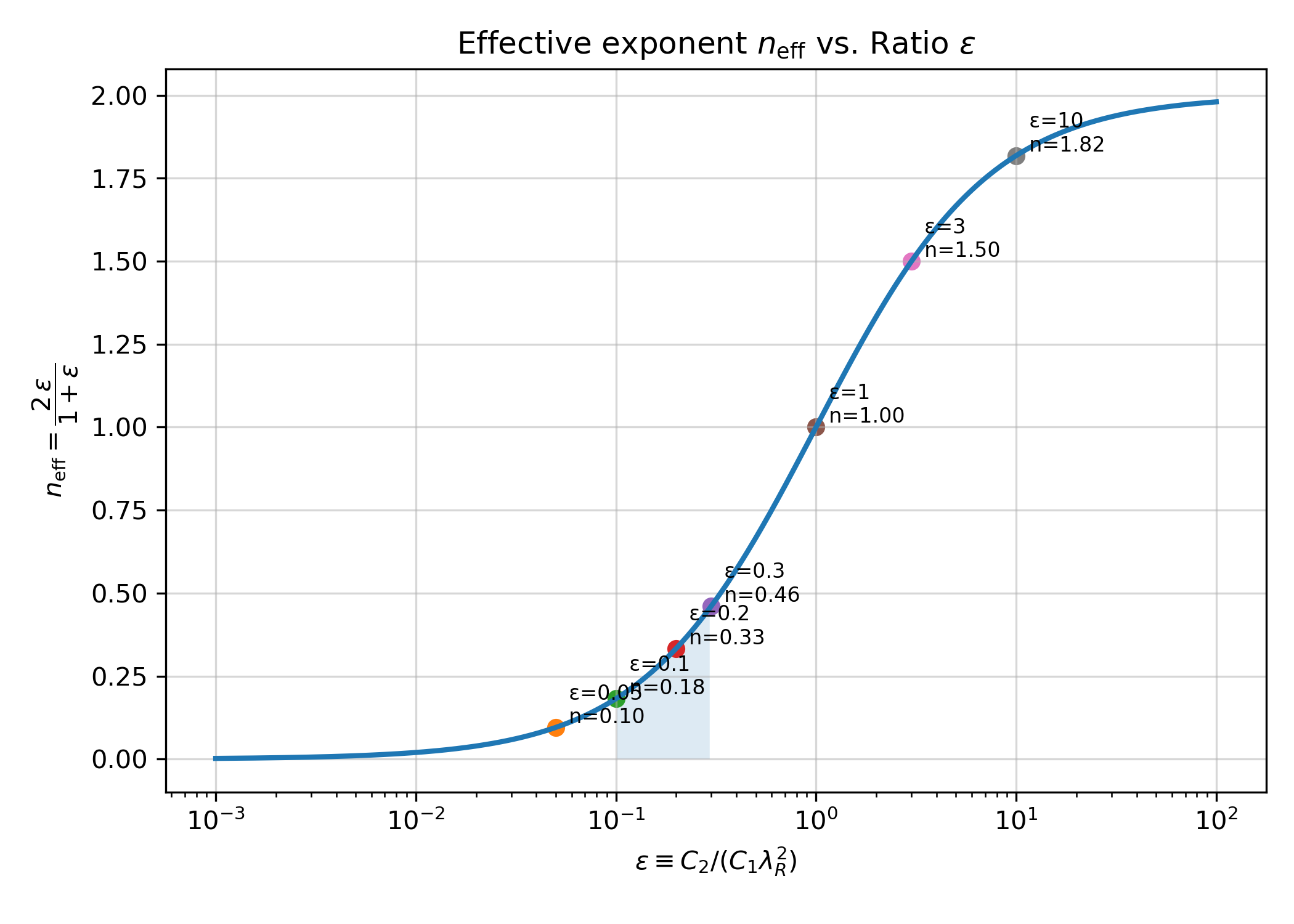}
    \caption{Effective exponent $n_{\rm eff}(\varepsilon)=2\varepsilon/(1+\varepsilon)$ 
as a function of the dimensionless ratio 
$\varepsilon=C_{2}/(C_{1}\lambda^{2})$. 
The curve interpolates smoothly between $n_{\rm eff}\to 0$ 
when the constant baseline dominates ($\varepsilon\ll 1$) 
and $n_{\rm eff}\to 2$ when the inverse-square term dominates 
($\varepsilon\gg 1$). 
The shaded interval $\varepsilon\sim0.1$--$0.3$ corresponds to the fitted 
values in our simulations, yielding $n_{\rm eff}\sim0.3$--$0.4$, consistent 
with the shallow slopes reported in Table~\ref{table:fitting_par}.}
    \label{fig:epsilon}
\end{figure}

Two immediate consequences follow. First, because the framework is linear in flux and dominated by transport, detailed AR configuration (no/tilt/morphology asymmetry) affects $R(\lambda_R)$ only through modest changes in $C_1$ and $C_2$, so the three cases collapse to nearly the same curve. Second, the latitude at which LQ and TQ have equal impact is set by $R(\lambda_R^\ast)=1$, i.e.,
\[
\lambda_R^\ast \;\text{ satisfies }\; C_1+\frac{C_2}{(\lambda_R^\ast)^2}=1,
\]
which in our data lies near $\lambda_R^\ast\approx 12^\circ$; for $\lambda_R<\lambda_R^\ast$ latitude quenching dominates ($R>1$), whereas for $\lambda_R>\lambda_R^\ast$ tilt quenching dominates ($R<1$). If the Joy’s-law coupling term were suppressed (effectively $C_1\!\to\!0$), the ratio would revert toward a pure $1/\lambda_R^2$ dependence, closer to the standard SFT expectation. Figure~\ref{fig:epsilon} illustrates the dependence of $n_{\mathrm{eff}}$ on 
$\varepsilon$, confirming the shallow slopes  ($n_{\mathrm{eff}}\!\approx\!0.3$\--–$0.4$) derived in our fits.

\section{Discussion}\label{sect:disc.}

The algebraic approach reproduces the key behaviour of polar field generation seen in detailed surface flux-transport simulations. When nonlinear feedbacks are turned off, the final axial dipole at cycle minimum scales almost linearly with the cycle amplitude and with the dynamo effectivity range $\lambda_R$; this is expected since the polar field simply accumulates the contributions from each emerged bipole. Introducing Tilt quenching (TQ) alone already produces a noticeable nonlinearity: stronger cycles tend to have smaller Joy’s-law tilts, so each active region contributes less to the axial dipole.  This is consistent with observations of an anti-correlation between cycle strength and mean active-region tilt. Similarly, latitude quenching (LQ) by itself suppresses dipole growth in strong cycles, because more high-latitude emergences limit the cancellation of leading flux across the equator.  Crucially, when both TQ and LQ are applied simultaneously, the dipole amplitude exhibits a pronounced saturation (a “ceiling” effect). In practice, beyond a threshold cycle amplitude (roughly twice the reference value in our experiments), further increases in emergence strength yield diminishing returns in the net polar field. In other words, strong cycles become progressively self-limiting. This combined feedback provides a natural mechanism to regulate the solar dynamo: it prevents runaway growth of the polar field and helps explain why the Sun avoids unbounded cycle growth.

We quantify the balance between LQ and TQ by the ratio $R=\mathrm{dev}_{\mathrm{LQ}}/\mathrm{dev}_{\mathrm{TQ}}$. This ratio falls smoothly with increasing $\lambda_R$, following Equation~\ref{eq:fit}. Thus, in advection-dominated regimes (small $\lambda_R$), LQ provides the stronger dipole damping ($R$ $\ge 1$), whereas in diffusion-dominated regimes (large $\lambda_R$) TQ becomes dominant. Our results match analytic expectations and prior SFT-model studies: \cite{talafha2022role} found the same inverse-square dependence of the LQ/TQ balance on $\lambda_R$. In summary, we find that LQ dominates the nonlinear feedback only at low $\lambda_R$, while TQ dominates at high $\lambda_R$.

We also examined how active-region (AR) asymmetries affect dipole generation. Adding random tilt scatter (a small offset between the tilt angles of the bipolar polarities) introduces extra variability but does not qualitatively change the nonlinear trends: the suppression ratio curve retains its form and the transition point in $\lambda_R$ remains nearly the same. 

{By contrast, imposing realistic morphological asymmetry in the source leads to a smaller axial dipole, an effect that is most pronounced under LQ and LQTQ. In our algebraic framework, this is not a property of diffusion; rather, it follows from how the latitude–weighting operators act on the instantaneous source profile: enlarging the following polarity with the same amount of flux spreads its flux over latitude and reduces its polar–weighted contribution, while the more compact leading polarity contributes relatively more to the equatorial–weighted cancellation proxy. The net result is a weaker signed flux imbalance feeding the poles and hence a suppressed dipole. This behaviour is consistent with earlier studies: \citet{iijima2019effect} showed that leading–following morphological asymmetry can substantially reduce a BMR’s polar–field contribution (even reversing the expected sign in extreme cases), and \citet{wang2021algebraic} reported that algebraic/symmetric–BMR models systematically overestimate the dipole moments of real, asymmetric active regions.
}

Our findings fit well with previous theoretical and observational work. \cite{jiang2020nonlinear} investigated the combined effect of LQ and TQ using the SFT model and found that these two mechanisms contribute comparably to the poloidal source nonlinearity, leading to a saturation of the net dipole in strong cycles. This matches our conclusion using the algebraic approach that both mechanisms together impose a ceiling on polar field strength. Observationally, the strong anticorrelation between cycle strength and average tilt  \citep{dasi2010sunspot} lends empirical support to tilt quenching as a significant feedback.  Likewise, recent surface flux transport models have begun to include latitude-dependent quenching (as noted by \cite{petrovay2020solar} and explored by \cite{jiang2020nonlinear}), underscoring that both effects are important in realistic solar conditions.

These nonlinear saturation effects have important implications for solar cycle predictions.  The fact that very strong cycles produce only modest increases in the polar dipole provides a simple physical explanation for the Sun’s observed self-limiting behaviour.  It also suggests that the solar dynamo has a relatively short memory: if a large cycle is strongly quenched, its contribution to the next cycle’s seed field is reduced, weakening inter-cycle correlations.  

Our model has some important limitations to bear in mind: The algebraic source formulation is highly idealised; the meridional flow and diffusion are combined into a single parameter $\lambda_R$, and we assumed fixed Joy’s-law tilts and Gaussian emergence profiles. In the real Sun, flow speeds, diffusivity, and AR properties vary with latitude and phase of the cycle, and the Joy’s-law coefficient itself may evolve. Our simplification ignores these variations.

Future work can build on this framework in several ways. A natural next step is to calibrate the algebraic approach against solar observations and use data assimilation or physics-informed machine learning to predict Solar cycles.

\section{Conclusion}\label{sect:concl}

Our algebraic experiments confirm that both Latitude Quenching (LQ) and Tilt Quenching (TQ) individually weaken the axial dipole in strong cycles; acting together, they produce a clear saturation (a ``ceiling'') of polar field growth with increasing cycle amplitude. This supports the view that the solar cycle is intrinsically self-regulating: even very strong cycles do not yield proportionally stronger polar fields.

A central quantitative result is the behaviour of $R(\lambda_R)$. Using ensembles of independent realisations at each $\lambda_R$ and true $\chi^2$ fits with the ensemble $1\sigma$ scatter as absolute weights, we find that $R(\lambda_R)$ decreases monotonically and crosses unity at $\lambda_R\simeq 12^\circ$, marking a transition from LQ dominance at low $\lambda_R$ to TQ dominance at high $\lambda_R$.

Over the range $8^\circ\!\le\!\lambda_R\!\le\!20^\circ$, the data are well described by an \emph{offset power law}
\[
R(\lambda_R)=c_1+\frac{c_2}{\lambda_R^{\,n}},
\]
with a shallow exponent $n\simeq 0.36\pm 0.04$ (free-$n$ fits per case). In contrast, forcing a single ``SFT-style'' curve with $n=2$ to all points is statistically disfavoured for our parameter set (very large reduced $\chi^2$), indicating that the oft-assumed $1/\lambda_R^{2}$ scaling is not universal but depends on transport.

We also tested symmetric, tilt-asymmetric, and morphology-asymmetric prescriptions for active regions. Within uncertainties, all three collapse onto essentially the same $R(\lambda_R)$ curve: asymmetries introduce stochastic scatter (with a slightly higher reduced $\chi^2$ for the morphology-asymmetric case) but do not materially shift the trend or the transition scale. This insensitivity underscores that, for fixed transport parameters, the LQ–TQ balance is controlled primarily by $\lambda_R$ rather than by detailed AR configuration.

Taken together, these results establish the algebraic method as a fast, informative complement to full surface-flux-transport modelling for probing nonlinear saturation in the Babcock–Leighton dynamo. Calibrating $\lambda_R$ to observed transport and employing ensemble-weighted ($1\sigma$) $\chi^2$ fits provides a transparent pathway to quantify when and how LQ or TQ sets the ceiling on polar-field amplification—and thus on cycle-to-cycle variability.

%
\begin{acks}
The author gratefully acknowledges Dr Jie Jiang for insightful discussions on solar cycle modelling and constructive feedback that greatly improved the quality and clarity of this work. The author also thanks the Sharjah Academy for Astronomy, Space Sciences and Technology (SAASST) at the University of Sharjah for providing essential resources and infrastructure that supported this study. {Finally, the author also thanks the anonymous reviewer for their constructive comments and suggestions, which helped improve the clarity and rigour of the manuscript.}

\end{acks}

%
%
\begin{fundinginformation}
This research received no external funding.
\end{fundinginformation}
\begin{dataavailability}
The Python code used to implement the algebraic surface flux transport simulations, including the nonlinear quenching mechanisms, is openly available at \url{https://github.com/mtalafha90/Algebraic_Nonlin_SFT}.

\end{dataavailability}
%
%

%
%
\bibliographystyle{spr-mp-sola}
\bibliography{references}  

@article{lemerle2015coupled,
  title={A coupled 2$\times$ 2D Babcock--Leighton solar dynamo model. I. Surface magnetic flux evolution},
  author={Lemerle, Alexandre and Charbonneau, Paul and Carignan-Dugas, Arnaud},
  journal={The Astrophysical Journal},
  volume={810},
  number={1},
  pages={78},
  year={2015},
  publisher={IOP Publishing}
}

@article{schatten1978using,
  title={Using dynamo theory to predict the sunspot number during solar cycle 21},
  author={Schatten, Kenneth H and Scherrer, Philip H and Svalgaard, Leif and Wilcox, John M},
  journal={Geophysical Research Letters},
  volume={5},
  number={5},
  pages={411--414},
  year={1978},
  publisher={Wiley Online Library}
}

@article{wang1991magnetic,
  title={Magnetic flux transport and the sun's dipole moment-New twists to the Babcock-Leighton model},
  author={Wang, Y-M and Sheeley Jr, NR},
  journal={Astrophysical Journal, Part 1 (ISSN 0004-637X), vol. 375, July 10, 1991, p. 761-770. US Navy-supported research.},
  volume={375},
  pages={761--770},
  year={1991}
}

@article{wang1991new,
  title={A new solar cycle model including meridional circulation},
  author={Wang, Y-M and Sheeley Jr, NR and Nash, AG},
  journal={Astrophysical Journal, Part 1 (ISSN 0004-637X), vol. 383, Dec. 10, 1991, p. 431-442. Research supported by US Navy.},
  volume={383},
  pages={431--442},
  year={1991}
}

@article{jiao2021sunspot,
  title={Sunspot tilt angles revisited: dependence on the solar cycle strength},
  author={Jiao, Qirong and Jiang, Jie and Wang, Zi-Fan},
  journal={Astronomy \& Astrophysics},
  volume={653},
  pages={A27},
  year={2021},
  publisher={EDP Sciences}
}

@article{jiang2014effects,
  title={Effects of the scatter in sunspot group tilt angles on the large-scale magnetic field at the solar surface},
  author={Jiang, J and Cameron, RH and Sch{\"u}ssler, M},
  journal={The Astrophysical Journal},
  volume={791},
  number={1},
  pages={5},
  year={2014},
  publisher={IOP Publishing}
}

@article{petrovay2020solar,
	title={Solar cycle prediction},
	author={Petrovay, Krist{\'o}f},
	journal={Living Reviews in Solar Physics},
	volume={17},
	number={1},
	pages={2},
	year={2020},
	publisher={Springer}
}

@article{jiang2018predictability,
	title={Predictability of the solar cycle over one cycle},
	author={Jiang, Jie and Wang, Jing-Xiu and Jiao, Qi-Rong and Cao, Jin-Bin},
	journal={The Astrophysical Journal},
	volume={863},
	number={2},
	pages={159},
	year={2018},
	publisher={American Astronomical Society}
}

@article{petrovay2019optimization,
  title={Optimization of surface flux transport models for the solar polar magnetic field},
  author={Petrovay, K and Talafha, M},
  journal={Astronomy \& Astrophysics},
  volume={632},
  pages={A87},
  year={2019},
  publisher={EDP Sciences}
}

@article{petrovay2020towards,
  title={Towards an algebraic method of solar cycle prediction-I. Calculating the ultimate dipole contributions of individual active regions},
  author={Petrovay, Krist{\'o}f and Nagy, Melinda and Yeates, Anthony R},
  journal={Journal of Space Weather and Space Climate},
  volume={10},
  pages={50},
  year={2020},
  publisher={EDP Sciences}
}

@article{iijima2017improvement,
  title={Improvement of solar-cycle prediction: plateau of solar axial dipole moment},
  author={Iijima, H and Hotta, H and Imada, S and Kusano, K and Shiota, D},
  journal={Astronomy \& Astrophysics},
  volume={607},
  pages={L2},
  year={2017},
  publisher={EDP Sciences}
}

@article{cameron2010surface,
  title={Surface flux transport modeling for solar cycles 15--21: effects of cycle-dependent tilt angles of sunspot groups},
  author={Cameron, RH and Jiang, J and Schmitt, D and Sch{\"u}ssler, M},
  journal={The Astrophysical Journal},
  volume={719},
  number={1},
  pages={264},
  year={2010},
  publisher={IOP Publishing}
}

@article{dasi2010sunspot,
  title={Sunspot group tilt angles and the strength of the solar cycle},
  author={Dasi-Espuig, Maria and Solanki, Sami K and Krivova, Natasha A and Cameron, R and Pe{\~n}uela, Tania},
  journal={Astronomy \& Astrophysics},
  volume={518},
  pages={A7},
  year={2010},
  publisher={EDP Sciences}
}

@article{d1993theoretical,
  title={A theoretical model for tilts of bipolar magnetic regions},
  author={D'silva, S and Choudhuri, AR},
  journal={Astronomy and Astrophysics, Vol. 272, p. 621 (1993)},
  volume={272},
  pages={621},
  year={1993}
}

@article{longcope1996effects,
  title={The effects of convection zone turbulence on the tilt angles of magnetic bipoles},
  author={Longcope, DW and Fisher, GH},
  journal={Astrophysical Journal v. 458, p. 380},
  volume={458},
  pages={380},
  year={1996}
}

@article{talafha2025effect,
  title={Effect of Nonlinear Surface Inflows into Activity Belts on Solar Cycle Modulation},
  author={Talafha, Mohammed H and Petrovay, Krist{\'o}f and Opitz, Andrea},
  journal={Solar Physics},
  volume={300},
  number={5},
  pages={1--20},
  year={2025},
  publisher={Springer}
}

@article{iijima2019effect,
  title={Effect of morphological asymmetry between leading and following sunspots on the prediction of solar cycle activity},
  author={Iijima, H and Hotta, H and Imada, S},
  journal={The Astrophysical Journal},
  volume={883},
  number={1},
  pages={24},
  year={2019},
  publisher={IOP publishing}
}

@article{jiang2014magnetic,
  title={Magnetic flux transport at the solar surface},
  author={Jiang, J and Hathaway, DH and Cameron, RH and Solanki, Sami K and Gizon, Laurent and Upton, L},
  journal={Space Science Reviews},
  volume={186},
  pages={491--523},
  year={2014},
  publisher={Springer}
}

@article{jiang2011solar,
  title={The solar magnetic field since 1700-I. Characteristics of sunspot group emergence and reconstruction of the butterfly diagram},
  author={Jiang, Jie and Cameron, Robert H and Schmitt, Dieter and Schuessler, Manfred},
  journal={Astronomy \& Astrophysics},
  volume={528},
  pages={A82},
  year={2011},
  publisher={EDP Sciences}
}

@article{wang2021algebraic,
  title={Algebraic quantification of an active region contribution to the solar cycle},
  author={Wang, Zi-Fan and Jiang, Jie and Wang, Jing-Xiu},
  journal={Astronomy \& Astrophysics},
  volume={650},
  pages={A87},
  year={2021},
  publisher={EDP Sciences}
}

@article{baumann2004evolution,
  title={Evolution of the large-scale magnetic field on the solar surface: A parameter study},
  author={Baumann, I and Schmitt, D and Sch{\"u}ssler, M and Solanki, SK},
  journal={Astronomy \& Astrophysics},
  volume={426},
  number={3},
  pages={1075--1091},
  year={2004},
  publisher={EDP Sciences}
}

@article{jiang2016drives,
  title={What drives the solar magnetic cycle?},
  author={Jiang, J and Wang, JX and Zhang, JH and others},
  journal={Chin Sci Bull},
  volume={61},
  pages={2973--2985},
  year={2016}
}

@article{sekii2015recent,
  title={Recent discoveries of structures and physical processes in local helioseismology},
  author={Sekii, T and Shibahashi, H},
  journal={Extraterrestrial Seismology},
  pages={180--188},
  year={2015}
}

@inproceedings{nandy2004meridional,
  title={Meridional circulation and the solar magnetic cycle},
  author={Nandy, Dibyendu},
  booktitle={SOHO 14 Helio-and Asteroseismology: Towards a Golden Future},
  volume={559},
  pages={241},
  year={2004}
}

@article{choudhuri2018flux,
  title={Flux transport dynamo: From modelling irregularities to making predictions},
  author={Choudhuri, Arnab Rai},
  journal={Journal of Atmospheric and Solar-Terrestrial Physics},
  volume={176},
  pages={5--9},
  year={2018},
  publisher={Elsevier}
}

@article{kitchatinov2012solar,
  title={Solar dynamo model with nonlocal alpha-effect and diamagnetic pumping},
  author={Kitchatinov, LL and Olemskoy, SV},
  journal={Proceedings of the International Astronomical Union},
  volume={8},
  number={S294},
  pages={429--430},
  year={2012},
  publisher={Cambridge University Press}
}

@article{jha2020magnetic,
  title={Magnetic field dependence of bipolar magnetic region tilts on the Sun: indication of tilt quenching},
  author={Jha, Bibhuti Kumar and Karak, Bidya Binay and Mandal, Sudip and Banerjee, Dipankar},
  journal={The Astrophysical Journal Letters},
  volume={889},
  number={1},
  pages={L19},
  year={2020},
  publisher={IOP Publishing}
}

@article{choudhuri2023emergence,
  title={The emergence and growth of the flux transport dynamo model of the sunspot cycle},
  author={Choudhuri, Arnab Rai},
  journal={Reviews of Modern Plasma Physics},
  volume={7},
  number={1},
  pages={18},
  year={2023},
  publisher={Springer}
}

@article{pal2024algebraic,
  title={Algebraic quantification of the contribution of active regions to the Sun’s dipole moment: applications to century-scale polar field estimates and solar cycle forecasting},
  author={Pal, Shaonwita and Nandy, Dibyendu},
  journal={Monthly Notices of the Royal Astronomical Society},
  volume={531},
  number={1},
  pages={1546--1553},
  year={2024},
  publisher={Oxford University Press}
}

@article{hazra2016proposed,
  title={A proposed paradigm for solar cycle dynamics mediated via turbulent pumping of magnetic flux in Babcock--Leighton-type solar dynamos},
  author={Hazra, Soumitra and Nandy, Dibyendu},
  journal={The Astrophysical Journal},
  volume={832},
  number={1},
  pages={9},
  year={2016},
  publisher={IOP Publishing}
}

@article{karak2020dynamo,
  title={Dynamo saturation through the latitudinal variation of bipolar magnetic regions in the Sun},
  author={Karak, Bidya Binay},
  journal={The Astrophysical Journal Letters},
  volume={901},
  number={2},
  pages={L35},
  year={2020},
  publisher={IOP Publishing}
}

@article{martin2017inflows,
  title={Inflows towards active regions and the modulation of the solar cycle: A parameter study},
  author={Martin-Belda, David and Cameron, Robert H},
  journal={Astronomy \& Astrophysics},
  volume={597},
  pages={A21},
  year={2017},
  publisher={EDP Sciences}
}

@article{bhowmik2023physical,
  title={Physical models for solar cycle predictions},
  author={Bhowmik, Prantika and Jiang, Jie and Upton, Lisa and Lemerle, Alexandre and Nandy, Dibyendu},
  journal={Space Science Reviews},
  volume={219},
  number={5},
  pages={40},
  year={2023},
  publisher={Springer}
}

@article{karak2018recovery,
  title={Recovery from Maunder-like grand minima in a Babcock--Leighton solar dynamo model},
  author={Karak, Bidya Binay and Miesch, Mark},
  journal={The Astrophysical Journal Letters},
  volume={860},
  number={2},
  pages={L26},
  year={2018},
  publisher={IOP Publishing}
}

@article{jiang2020nonlinear,
  title={Nonlinear mechanisms that regulate the solar cycle amplitude},
  author={Jiang, Jie},
  journal={The Astrophysical Journal},
  volume={900},
  number={1},
  pages={19},
  year={2020},
  publisher={IOP Publishing}
}

@article{karak2017solar,
  title={Solar cycle variability induced by tilt angle scatter in a Babcock--Leighton solar dynamo model},
  author={Karak, Bidya Binay and Miesch, Mark},
  journal={The Astrophysical Journal},
  volume={847},
  number={1},
  pages={69},
  year={2017},
  publisher={IOP Publishing}
}

@article{cameron2012strengths,
  title={Are the strengths of solar cycles determined by converging flows towards the activity belts?},
  author={Cameron, RH and Sch{\"u}ssler, M},
  journal={Astronomy \& Astrophysics},
  volume={548},
  pages={A57},
  year={2012},
  publisher={EDP Sciences}
}

@article{nagy2020impact,
  title={Impact of nonlinear surface inflows into activity belts on the solar dynamo},
  author={Nagy, Melinda and Lemerle, Alexandre and Charbonneau, Paul},
  journal={Journal of Space Weather and Space Climate},
  volume={10},
  pages={62},
  year={2020},
  publisher={EDP Sciences}
}

@article{sood2014detailed,
  title={Detailed mathematical and numerical analysis of a dynamo model},
  author={Sood, Aditi and Kim, Eun-jin},
  journal={Astronomy \& Astrophysics},
  volume={563},
  pages={A100},
  year={2014},
  publisher={EDP Sciences}
}

@article{kleeorin2023nonlinear,
  title={Nonlinear Mean-Field Dynamos With Magnetic Helicity Transport and Solar Activity: Sunspot Number and Tilt},
  author={Kleeorin, Nathan and Kuzanyan, Kirill and Rogachevskii, Igor and Safiullin, Nikolai},
  journal={Helicities in Geophysics, Astrophysics and Beyond},
  pages={217--240},
  year={2023},
  publisher={Wiley Online Library}
}

@article{yeates2025latitude,
  title={Latitude Quenching Nonlinearity in the Solar Dynamo},
  author={Yeates, Anthony R and Bertello, Luca and Pevtsov, Alexander A and Pevtsov, Alexei A},
  journal={The Astrophysical Journal},
  volume={978},
  number={2},
  pages={147},
  year={2025},
  publisher={IOP Publishing}
}

@article{talafha2022role,
  title={Role of observable nonlinearities in solar cycle modulation},
  author={Talafha, M and Nagy, M and Lemerle, A and Petrovay, K},
  journal={Astronomy \& Astrophysics},
  volume={660},
  pages={A92},
  year={2022},
  publisher={EDP Sciences}
}
%
%
%
%

\end{document}